\documentclass{article}
\begin{document}

\newcommand{\mb}[1]{{\mbox{\boldmath{$#1$}}}}

\thispagestyle{plain} \setcounter{page}{1}

\input amssym.tex

\title{Infinite loop
superalgebras of the Dirac theory on
the Euclidean Taub-NUT space}

\author{Ion I. Cot\u aescu \thanks{E-mail:~~~cota@physics.uvt.ro}\\
{\small \it West University of Timi\c soara,}\\
       {\small \it V. P\^ arvan Ave. 4, RO-300223 Timi\c soara, Romania}
\and
Mihai Visinescu \thanks{E-mail:~~~mvisin@theory.nipne.ro}\\
{\small \it Department of Theoretical Physics,}\\
{\small \it National Institute for Physics and Nuclear Engineering,}\\
{\small \it P.O.Box M.G.-6, Magurele, Bucharest, Romania}}
\date{}

\maketitle

\begin{abstract}

The Dirac theory in the Euclidean Taub-NUT space
gives rise to a large collection of
conserved operators associated to genuine or hidden symmetries.
They are involved in interesting algebraic structures as dynamical
algebras or even infinite-dimensional algebras or superalgebras.
One presents here the infinite-dimensional superalgebra specific
to the Dirac theory in manifolds carrying the Gross-Perry-Sorkin
monopole. It is shown that there exists an infinite-dimensional
superalgebra that can be seen as a twisted loop superalgebra.

Pacs 04.62.+v

\end{abstract}

\section{Introduction}

In the quantum physics on curved space-times an interesting
problem is to find the algebras of operators that commute with the
field equation. In general these operators have to be the
generators of isometries \cite{CML,ES} or special operators
associated with more subtle hidden symmetries that can occur in
association with some supersymmetries \cite{MaCa}.

We mention that there are two generalization of the Killing  vectors which
become of interest in physics, namely the St\" ackel-Killing (S-K) tensors and
the Killing-Yano (K-Y) tensors. A symmetric tensor field $K_{\mu_1...\mu_r}$ is
called a S-K tensor  of valence $r$ if $ K_{(\mu_1...\mu_r;\lambda)} = 0$. The
usual Killing (K) vectors correspond to valence $r=1$ while the hidden
symmetries are encapsulated in S-K tensors of valence $r>1$.  A  tensor
$f_{\mu_1...\mu_r}$ is called a  K-Y tensor of valence $r$ if it is totally
antisymmetric and it satisfies the equation \cite{Y} $
f_{\mu_1...(\mu_r;\lambda)} = 0$.

An  example of a background presenting all these types of symmetries and
supesymmetries is the space-time of the Gross-Perry-Sorkin (GPS) monopole
defined as the Euclidean Taub-NUT space with the time trivially added
\cite{GPS}. The space part is known to be a hyper-K\" ahler manifold possessing
three covariantly constant Killing-Yano (K-Y) tensors with real-valued
components which constitute a hypercomplex structure. This generates a ${\cal
N}=4$ superalgebra of Dirac-type operators \cite{CV2}, in a similar way as in
pseudo-classical spinning models \cite{VV1,VV2}. The Euclidean Taub-NUT space
has, in addition, a non-covariantly constant K-Y tensor related to its specific
hidden symmetry \cite{GR,GR1,JWH,VV1,VV2} giving a conserved Runge-Lenz type
operator \cite{GH,GM,GR,FH,CFH}. In the Dirac theory the corresponding operator
can be constructed
with the help of the Dirac-type operators produced by the four K-Y tensors of
this space \cite{CV3,CV4}. Thus one obtains a rich algebra of conserved
observables \cite{CV5} that offers many possibilities of choosing sets of
commuting operators for defining quantum modes \cite{CV1,CV2}. On the other
hand, hereby one can select dynamical algebras typical for the Kepler
problems \cite{CV4,CV5}, or even interesting infinite-dimensional algebras or
superalgebras \cite{NOVA}.
Similar problems appear in the study of chiral supersymmetry for spin fields
in self-dual backgrounds \cite{FHR}.

Our main objective here is to present the content of the operator algebra of
the Dirac theory on Euclidean Taub-NUT space showing that this can be seen as a
twisted loop superalgebra in the sense of \cite{DD}. To this end we review our
previous results paying more attention to the Casimir operators involved in our
construction. Thus we define a new infinite-dimensional superalgebra, different
from that given in \cite{NOVA}, that leads to a twisted loop superalgebra in a
natural way.

We start in section 2 presenting the main features of the Euclidean Taub-NUT
geometry and the operators of the scalar quantum theory pointing out some
useful algebraic properties. The next section is devoted to the relationships
among the Pauli and Dirac conserved operators that are given in section 4. In
section 5 we construct our improved version of infinite-dimensional
superalgebra showing in section 6 that this is a twisted loop superalgebra.

\section{Euclidean Taub-NUT space}

The manifold of the GPS monopole, denoted from now by ${\frak M}$, is a
5-dimensional Kaluza-Klein space-time whose space part is the Euclidean
Taub-NUT space. There are static charts with Cartesian coordinates $x^{\mu}$
($\mu, \nu,...=0,1,2,3,4$) where the time is $t=x^0$, $x^{i}$ ($i,j,...=1,2,3$)
are the {\em physical} Cartesian space coordinates while $x^4$ is the Cartesian
extra-coordinate. Taking the metric of the flat model $\eta=(-1,1,1,1,1)$  we
can use the three-dimensional vector notations, $\vec{x}=(x^1,x^2,x^3)$,
$r=|\vec{x}|$  and $dl^{2}=d\vec{x}\cdot d\vec{x}$, for writing the GPS line
element
\begin{equation}\label{3(met)}
ds^2=-dt^2+\frac{1}{V(r)}dl^2 + V(r)[dx^4+A^{em}_i(\vec{x})dx^i]^2
\,,
\end{equation}
defined by the specific functions
\begin{equation}\label{3(tn)}
\frac{1}{V}=1+\frac{\mu}{r}\,,\quad
A^{em}_{1}=-\frac{\mu}{r}\frac{x^{2}}{r+x^{3}}\,, \quad
A^{em}_{2}=\frac{\mu}{r}\frac{x^{1}}{r+x^{3}}\,,\quad
A^{em}_{3}=0\,.
\end{equation}
The real number $\mu$ is a parameter of the theory. If one
interprets $\vec{A}^{em}$ as the vector potential (or gauge field)
it results  the magnetic field with central symmetry
\begin{equation}
\vec{B}^{em}\,=\mu\frac{\vec{x}}{r^3}\,.
\end{equation}

The Taub-NUT geometry possesses a special type of isometries which form the
isometry group $I({\frak M})=T(1)_t\otimes SO(3)\otimes U(1)_4$ constituted by
time translations, space rotations and $U(1)$ transformations of
extra-coordinate. The universal covering group of $I({\frak M})$ is the
external symmetry group (in the sense of \cite{ES}) $S({\frak M})=T(1)_t\otimes
SU(2)\otimes U(1)_4$. In Ref. \cite{CV8} we pointed out that these isometries
combine space transformations with  gauge ones in a non-trivial manner
generating non-linear transformations of the coordinate $x^4$ under rotations.
Fortunately, the complications due to this phenomenon can be avoided in a
special gauge where the symmetry under rotations becomes {\em global}. In the
Cartesian charts this gauge is given by the gauge fields $\hat e^{\hat\alpha}$
and $e_{\hat\alpha}$ having the non-vanishing components \cite{BuCh}
\begin{eqnarray}
&&\hat e^0_0=1\,,\quad \hat
e^{i}_{j}=\frac{1}{\sqrt{V}}\,\delta_{ij}\,, \quad \hat
e^{4}_{i}=\sqrt{V}A^{em}_{i}\,, \quad
\hat e^{4}_{4}=\sqrt{V}\,, \nonumber\\
&&e^0_0=1\,,\quad e^{i}_{j}=\sqrt{V}\delta_{ij}\,,\quad
e^{4}_{i}=-\sqrt{V}A^{em}_{i}\,,\quad e^{4}_{4}=\frac{1}{\sqrt{V}}
\,.\label{ehe}
\end{eqnarray}

In this context one can correctly define $P_4=-i\partial_4$ and
the three-di\-men\-sio\-nal physical momentum $\vec{P}$ whose components
(in the mentioned local frames) are ${P}_i=
-i(\partial_i-{A}^{em}_i\partial_{4})$. 
Moreover, the angular momentum can be written in covariant form as
\begin{equation}\label{3(angmom)}
\vec{L}\,=\,\vec{x}\times\vec{P}-\mu\frac{\vec{x}}{r}P_4\,.
\end{equation}
These operators obey
$[P_{i},P_{j}]=i\varepsilon_{ijk}B_{k}^{em}P_{4}$,
$[P_{i},P_{4}]=0$ and $[L_i, \,P_j]=i\varepsilon_{ijk}P_{k}$ which
indicate that $\vec{P}$  behaves as a vector under rotations.
The scalar quantum mechanics in GPS geometry \cite{CFH} is based
on the Schr\" o\-din\-ger or Klein-Gordon equations involving the
static operator
\begin{equation}
\Delta=V{\vec{P}\,}^{2}+\frac{1}{V}{P_{4}}^{2} \,,
\end{equation}
which is either proportional with the Hamiltonian operator of the
Schr\" o\-din\-ger theory  or represents the static part of the
Klein-Gordon operator \cite{CV5}.

The space part of the manifold with GPS monopole is the Euclidean
Taub-NUT space which is a hyper-K\" ahler manifold possessing a
triplet of hypercomplex structures, ${\bf f}=\{
f^{(1)},\,f^{(2)},\, f^{(3)}\}$, defined as
\begin{equation}\label{trip}
f^{(i)}= f^{(i)}_{\hat\alpha \hat\beta} \hat e^{\hat\alpha}\land
\hat e^{\hat\beta}= 2\hat e^i\land \hat
e^4-\varepsilon_{ijk}\hat e^j\land \hat e^k \,,
\end{equation}
where the 1-forms $ \hat e^{\hat \alpha}= \hat e^{\hat
\alpha}_{\mu}dx^{\mu}$ are defined by the gauge fields
(\ref{ehe}). In addition, there exists a fourth K-Y tensor,
\begin{equation}\label{3fy}
f^Y = f^Y_{\hat\alpha \hat\beta}\hat e^{\hat\alpha}\land \hat
e^{\hat\beta}=
 \frac{x^i}{r}f^{(i)} +\frac{2 x^i}{\mu V}\varepsilon_{ijk}
\,{\hat e}^j\land \,{\hat e}^k\,,
\end{equation}
which is not covariantly constant. The presence of $f^Y$ is related to the
existence of the hidden symmetries of the Euclidean Taub-NUT geometry,
encapsulated in three non-trivial {S-K} tensors, $k_i^{\mu\nu}$. These are
interpreted as the components of the so-called Runge-Lenz vector of the
Euclidean Taub-NUT geodesics and can be expressed as symmetrized products of
K-Y tensors \cite{VV2,VV1}. The corresponding conserved vector operator,
\begin{equation}
\vec{K}\,=-\frac{1}{2}\nabla_{\mu}\vec{k}^{\mu\nu}\nabla_{\nu}=
\frac{1}{2}\left(\vec{P}\times \vec{L}- \vec{L}\times
\vec{P}\right)- \mu \frac{\vec{x}}{r}
\left(\frac{1}{2}\Delta-P_4^{2}\right)\,,
\end{equation}
play the same role as the Runge-Lenz vector operator in the usual
quantum mechanical Kepler problem \cite{CFH}.

This operator transforms as a vector under rotations such that one can write
the following complete system of commutation relations
\begin{eqnarray}
\left[ L_{i},\, L_{j} \right]& =& i \varepsilon_{ijk}\,L_{k}\,,\nonumber\\
\left[ L_{i},\, K_{j} \right]& =& i \varepsilon_{ijk}\,K_{k}\,,\label{LLKK} \\
\left[ K_{i},\, K_{j} \right]& =& i
\varepsilon_{ijk}\,L_{k}B^2\,,\nonumber
\end{eqnarray}
where $B^2 ={P_4}^2-\Delta$. The operators $L_i$ and $K_i$ commute
with $B$ since they commute with $\Delta$ and $P_4$. Moreover, it
is known \cite{FH} that the operators
\begin{eqnarray}
&&C_1=\vec{L}^2B^2+\vec{K}^2=\mu^2P_4^2B^2+\frac{\mu^2}{4}
\left(B^2+{P_4}^2\right)^2-B^2 \label{C1C2}\\
&&C_2=\vec{L}\cdot
\vec{K}=-\frac{\mu^2}{2}P_4(B^2+P_4^2)\,,\label{C1C21}
\end{eqnarray}
play the role of Casimir operators for the open algebra (\ref{LLKK}).  With
their help we can define the new Casimir operators,
\begin{equation}\label{CBCB}
C^{\pm}=C_1\pm 2B C_2 +B^2=\frac{\mu^2}{4}(P_4\mp B)^4=B^2(N\pm \mu P_4)^2\,.
\end{equation}
where $N$ is the operator whose eigenvalues are just the values of the
principal quantum number of the discrete energy spectra \cite{NOVA}.

We note that the algebra (\ref{LLKK}) does not close to a finite Lie algebra
because of the factor $B^2$ that affects the last commutation relation.
Nevertheless, one can obtain Lie algebras replacing the operators $P_4$ and
$\Delta$ by their eigenvalues $\hat q$ and respectively $E^2$. Then one can
replace $B^2$  by $\hat q^2-E^2$ and rescale the generators $K_i$. In this
manner one obtains three different dynamical algebras:  the $o(4)$ algebra for
the discrete energy spectrum in the domain $E<|\hat q|$, the $o(3,1)$ algebra
for continuous spectrum in the domain $E>|\hat q|$ and the $e(3)$ algebra
corresponding only to the ground energy of the continuous spectrum, $E=|\hat
q|$ \cite{GR,GR1,FH,CFH}.

\section{Conserved  Dirac and Pauli operators}

For building the Dirac theory we consider the Cartesian chart, the
usual four-dimensional space of the Dirac spinors, $\Psi$, and the
Dirac matrices $\gamma^{\hat\alpha}$, that  satisfy $\{
\gamma^{\hat\alpha},\, \gamma^{\hat\beta} \} =2\eta_{\hat\alpha
\hat\beta}$, in the following  representation
\begin{equation}\label{3(gammai)}
\gamma^i = -i \left(
\begin{array}{cc}
0&\sigma_i\\
-\sigma_i&0
\end{array}\right)\,,\quad
\gamma^4 = \left(
\begin{array}{cc}
0&{\bf 1}_2\\
{\bf 1}_2&0
\end{array}\right)\,,
\end{equation}
where $\sigma_i$ are the Pauli matrices. In addition, we take
$\gamma^0=i\gamma^1\gamma^2\gamma^3\gamma^4=i\,{\rm diag}({\bf 1}_2, -{\bf
1}_2)$. With these notations  the {\it standard} Dirac operator without
explicit mass term  reads $D =\gamma^{\alpha}\nabla_{\alpha}$  \cite{CV2,CV3}
giving the corresponding {\em massless} Hamiltonian operator \cite{CV2,CV4}
\begin{equation}\label{HH}
H =-i\gamma^0 D =\left(
\begin{array}{cc}
0&{\mb \alpha}^{*}\\
{\mb \alpha}&0
\end{array}\right)
=\left(
\begin{array}{cc}
0&V\pi^{*}\frac{1}{\sqrt{V}}\\
\sqrt{V}\pi&0
\end{array}\right)\,,
\end{equation}
where $\pi = {\sigma}_{P}-iV^{-1}P_{4}$ and $\pi^* =
{\sigma}_{P}+iV^{-1}P_{4}$ depending on
$\sigma_P=\vec{\sigma}\cdot\vec{P}$. These  operators obey
\begin{equation}\label{daapipi}
\Delta= {\mb \alpha}^{*}{\mb \alpha}=V\pi^{*}\pi\,.
\end{equation}
We specify that here the star superscript is a mere notation that does not
coincide with the Hermitian conjugation of the Pauli operators. The operator
$H$ is the central piece of the Dirac theory and has the remarkable property to
produce the {\em  same} energy spectrum as those given by the static
Klein-Gordon equation, $\Delta \phi=E^2 \phi$.

Here we focus on the {\em conserved} operators of the Dirac theory which {\em
commute} with $H$. We denote by ${\bf D}=\{ X\,|\, [X,H]=0\}$ the algebra of
the conserved Dirac  operators observing that they can be related to Pauli
operators commuting with $\Delta$ which form the algebra ${\bf P}=\{\hat X\,|\,
[\hat X, \Delta]=0 \}$ where we include the orbital operators having this
property. All these operators are considered as conserved operators in the
sense of the Klein-Gordon theory. Notice that the Pauli operators are
interesting here since they are involved in different versions of the dyon
theory \cite{DYON} (see also \cite{FHR}) which may be compared to our approach.

In Ref. \cite{CV5} we have demonstrated that for any conserved
Pauli operator $\hat X \in {\bf P}$ we can construct  the diagonal
Dirac operator
\begin{equation}
{\cal D}(\hat X)=\left(
\begin{array}{cc}
\hat X&0\\
0&{\mb \alpha}\hat X\Delta^{-1}{\mb \alpha}^{*}
\end{array}\right) \,,
\end{equation}
which is also conserved. Particularly, for $\hat X={\bf 1}_2$ we
obtain the projection operator
\begin{equation}\label{id}
I ={\cal D}({\bf 1}_2)=\left(
\begin{array}{cc}
{\bf 1}_2&0\\
0&{\mb \alpha}\Delta^{-1}{\mb \alpha}^{*}
\end{array}\right) \,,
\end{equation}
on the space $\Psi_D = I \Psi $ in which the eigenspinors $\psi_E$
of $H$ form a (generalized) basis. This projection operator splits
the algebra ${\bf D}={\bf D}_{0}\oplus {\bf D}_{1}$ in two
subspaces of the projections $XI\in {\bf D}_{0}$ and $X({\bf
1}-I)\in {\bf D}_{1}$ of all $X\in {\bf D}$. One can demonstrate
that the subalgebra ${\bf D}_1$ is an ideal in ${\bf D}$
\cite{CV5}.

Another type of conserved Dirac operators are the ${\cal Q}$-operators defined
in \cite{CV2} as
\begin{equation}
{\cal Q}(\hat X)=\left\{ H\,,\,\left(
\begin{array}{cc}
\hat X&0\\
0&0
\end{array}
\right) \right\} =\left(
\begin{array}{cc}
0&\hat X{\mb \alpha}^*\\
{\mb \alpha}\hat X&0
\end{array}
\right)\,,
\end{equation}
where $\hat X$ may be any Pauli operator. However, if $\hat X\in
{\bf P}$ then ${\cal Q}(\hat X)\in {\bf D}_{0}$ since $[{\cal
Q}(\hat X),H]=0$ and ${\cal Q}(\hat X)I= {\cal Q}(\hat X)$. If
$\hat X={\bf 1}_2$ we obtain just the Hamiltonian operator
$H={\cal Q}({\bf 1}_2)\in {\bf D}_0$. Consequently, the inverse of
$H$ with respect to $I$ can be represented as $H^{-1}={\cal
Q}(\Delta^{-1})$. The mappings ${\cal D} : {\bf P}\to {\bf D}_{0}$
and ${\cal Q} : {\bf P}\to {\bf D}_{0}$ are linear and have the
following algebraic properties
\begin{eqnarray}
{\cal D}(\hat X){\cal D}(\hat Y)&=&{\cal D}(\hat X \hat Y)\,,\\
{\cal Q}(\hat X){\cal Q}(\hat Y)&=&{\cal D}(\hat X \hat Y\Delta)\,,\\
{\cal D}(\hat X){\cal Q}(\hat Y)&=& {\cal Q}(\hat X){\cal D}(\hat
Y)={\cal Q}(\hat X \hat Y)\,,
\end{eqnarray}
for any  $\hat X,\, \hat Y \in {\bf P}$. Moreover, the relations
$[\gamma^0,\, {\cal D}(\hat X)]=0$ and $\{\gamma^0,\, {\cal
Q}(\hat X)\}\\= 0$ show us that, according to the usual terminology
\cite{TH}, ${\cal D}$ and $\gamma^0{\cal D}$ are {\em even} Dirac
operators while ${\cal Q}$ and $\gamma^0{\cal Q}$ are {\em odd}
ones. We note that there are many other odd or even operators
which do not have such forms.

Since $I$ is the projection operator on the space of the Dirac
spinors $\Psi_D$ we say that the projection $IXI \in {\bf D}_0$ of
any Dirac operator $X$ represents the {\em physical part} of $X$.
The physical part of any Dirac operator is conserved and can be
written in terms of ${\cal D}$ or ${\cal Q}$-operators
\cite{CV4,CV5}. The action of $X$ reduces thus to that of  Pauli
operators allowing us to rewrite the problems of the Dirac theory
in terms of Pauli operators \cite{CV3,CV4}.

Notice that the off-diagonal operators can be transformed at any time in
diagonal ones using the multiplication with $H$ or $H^{-1}$. For example,  $H$
itself which is off-diagonal is related to the diagonal operators $H^2={\cal
D}(\Delta)$ or  $I$. Thus each  Dirac operator from ${\bf D}$ can be brought in
a diagonal form  associated with an operator from ${\bf P}$.

\section{The operators of the Dirac theory}

Carter and McLenaghan showed that in the theory of Dirac fermions for any
isometry with K vector $R_\mu$ there is an appropriate operator \cite{CML}:
\begin{equation}
X_k = -i ( R^\mu \hat\nabla_\mu - \frac{1}{4} \gamma^\mu \gamma^\nu
R_{\mu;\nu})
\end{equation}
which commutes with the standard Dirac operator $D$. In this geometry among the
K vectors corresponding to the  $S({\frak M})$ generators only one is
time-like,  $i\partial_t$, generating time-translations. The other K vectors
are time-independent, giving rise to conserved operators.

In other respects, each K-Y tensor $f_{\mu\nu}$ produces a {\it non-standard}
Dirac operator of the form
\begin{equation}\label{df}
D_f = -i\gamma^\mu (f_\mu ^{~\nu}\hat\nabla_\nu  -
\frac{1}{6}\gamma^\nu
\gamma^\rho f_{\mu\nu;\rho})
\end{equation}
which {\em anticommutes} with the standard one, $D$. In the case of the
Euclidean Taub-NUT space Dirac-type operators are constructed from the K-Y
tensors of this metric.

The simplest operators of ${\bf D}$ which commute with $H$, $D$, and $\gamma^0$
are the generators of the spinor representation of $S({\frak M})$ carried by
the space $\Psi$. The expressions of these operators are strongly dependent on
the gauge fixing. For this reason one prefers the gauge (\ref{ehe}) where the
spinor fields transform {\em manifestly covariant} under isometries. In this
gauge, the rotation generators of the spinor representation of $S({\frak M})$
are just the standard components of the {\em total} angular momentum
\begin{equation}
{J}_i=L_i+S_i\,, \quad S_i=\textstyle{\frac{1}{2}\varepsilon_{ijk}S^{jk}
=\frac{1}{2}}{\rm diag}(\sigma_i,\sigma_i)\,,
\end{equation}
with point-independent spin operators  \cite{CV2}. In the same way one can show
that the $U(1)_4$ generator, $P_4$, does not get a spin term. Hence it results
that the spinor representation of $S({\frak M})$ is {\em reducible} being a sum
of two irreducible representations carried by spaces of two-dimensional Pauli
spinors where the components of the total angular momentum are  $\hat J_i=L_i
+\frac{1}{2}\sigma_i$. Moreover, the physical part of the total angular
momentum reads
\begin{equation}\label{JI} {\cal J}_{i}= J_i I= {\cal D}(\hat
J_{i})={\cal D}(L_{i}) + \textstyle\frac{1}{2} {\cal D}(\sigma_{i}) \,,
\end{equation}
where both the orbital and the spin terms  {\em separately}
commute with $H$ since $L_{i}$ and $\sigma_{i}$ commute with
$\Delta$.

The triplet ${\bf f}$ defined by Eq.  (\ref{trip}) gives rise to
the spin-like operators
\begin{equation}\label{3sl}
\Sigma^{(i)}=\frac{i}{4} f^{(i)}_{\hat{\alpha}\hat\beta}
\gamma^{\hat{\alpha}}\gamma^{\hat\beta}=\left(
\begin{array}{cc}
\sigma_i&0\\
0&0
\end{array}\right)\,,
\end{equation}
and produce the  Dirac-type operators
 \cite{CV2}
\begin{equation}\label{3Dto}
D^{(i)}= -f^{(i)}_{\mu,\nu}
\gamma^{\nu}\nabla^{\mu}=i[D,\,\Sigma^{(i)}]= -i\left(
\begin{array}{cc}
0&\sigma_i {\mb \alpha}^*\\
{\mb \alpha}\sigma_i&0
\end{array}\right)=-i {\cal Q}(\sigma_i) \,,
\end{equation}
which anticommute with $D$ and $\gamma^0$. The operators $D$ and $D^{(i)}$,
$i=1,2,3$, form the basis of the ${\cal N}=4$ superalgebra \cite{CV4}. In
current calculations, when one is not interested to exploit the ${\cal N}=4$
superalgebra, it is indicated to use the simpler operators
\begin{equation}
Q_i= i H^{-1}D^{(i)}=H^{-1}{\cal Q}(\sigma_{i})={\cal
D}(\sigma_{i}) \,,
\end{equation}
instead of $D^{(i)}$. However, in this case the fourth partner of
the operators $Q^i$ is rather trivial since this is just $I$.
Therefore, these operators form a representation of the quaternion
units (or of the algebra of Pauli matrices) with values in ${\bf
D}_{0}$,
\begin{equation}\label{Pau}
Q_{i}\, Q_{j}=\delta_{ij}I+i\varepsilon_{ijk} Q_{k}\,,
\end{equation}
producing an evident ${\cal N}=3$ superalgebra.

The corresponding Dirac-type operator of the last K-Y tensor,
$f^Y$ was obtained in  \cite{CV3}.  This has the form
\begin{equation}\label{dy1}
D^Y=-{\cal Q}(\sigma_r) +\frac{2i}{\mu\sqrt{V}}\left(
\begin{array}{cc}
0&\lambda\\
-\lambda&0
\end{array}
\right)\,,
\end{equation}
where the Pauli operators $\sigma_r=\vec{\sigma}\cdot \vec{x}/r$
and $\lambda=\sigma_{L}+{\bf 1_2}+\mu\sigma_{r}P_{4}$ have
suitable properties that  help one to find the equivalent forms
reported in \cite{CV3} and verify that $D^Y$ commutes with $H$ and
$P_{4}$ and anticommutes with ${D}$ and $\gamma^0$. Moreover, we
observe that the physical part of $D^Y$ can be put in the form
\begin{equation}
D^{Y}I={\cal Q}(\sigma^{Y}\Delta^{-1}) \,, \quad
\sigma^{Y}=\frac{2}{\mu}\left[\sigma_{K}+(\sigma_{L}+{\bf 1}_2)P_{4}\right]\,,
\end{equation}
where $\sigma^Y$ is a new  conserved Pauli operator  associated to
\begin{equation}
Q^Y=H D^{Y}=H D^{Y}I= {\cal D}(\sigma^{Y})\in {\bf D}_0.
\end{equation}
We note that the Pauli operators $\sigma_L=\vec{\sigma}\cdot
\vec{L}$ and $\sigma_K=\vec{\sigma}\cdot \vec{K}$ are conserved
and satisfy $\{\sigma_K,\,\sigma_L+{\bf 1}_2\}=2\vec{L}\cdot
\vec{K}$ and $\{\sigma_r,\,\sigma_L+{\bf 1}_2\}= -2\mu P_4$.

As in the case of the Klein-Gordon theory, we can define  the
components of the conserved Runge-Lenz operator of the Dirac
theory  \cite{CV3,CV4} giving directly their physical parts,
\begin{equation}\label{KI}
{\cal K}_{i}=\frac{\mu}{4}\{ Q^Y,\, Q_i\} + \frac{1}{2}({\cal B}-P_4)Q_i-{\cal
J}_i  P_4 \in {\bf D}_0\,,
\end{equation}
where ${\cal B}^{2}={P_{4}}^{2}I-H^2={\cal D}(B^2)$. Consequently, we can
express
\begin{equation}\label{Kas}
{\cal K}_{i}= {\cal D}(\hat K_{i})\,,\quad \hat K_{i}=K_{i}
+\frac{\sigma_{i}}{2}\,B\in {\bf P} \,.
\end{equation}
The operators ${\cal J}_i$ and ${\cal K}_i$ are involved in the
following system of commutation relations
\begin{eqnarray}
\left[ {\cal J}_{i},\, {\cal J}_{j}\right]&=&i\varepsilon
_{ijk}{\cal J}_{k}
\,,\nonumber\\
\left[ {\cal J}_{i},\, {\cal K}_{j}\right]&=&i\varepsilon
_{ijk}{\cal K}_{k}
\,,\label{algJK}\\
\left[ {\cal K}_{i},\, {\cal K}_{j}\right]&=&i\varepsilon
_{ijk}{\cal J}_{k} {\cal B}^{2}\,,\nonumber
\end{eqnarray}
and commute with the operators $Q_i$ as \cite{CV5},
\begin{equation}\label{QJK}
\left[ {\cal J}_{i},\, {Q}_{j}\right]=i\varepsilon _{ijk}{Q}_{k}
\,,\qquad \left[ {\cal K}_{i},\, {Q}_{j}\right]=i\varepsilon
_{ijk}{Q}_{k} {\cal B}\,.
\end{equation}
The algebra (\ref{algJK}) does not close as a Lie algebra because
of the factor ${\cal B}^2$. The dynamical algebras of the Dirac
theory have to be obtained as in the scalar case by replacing this
operator with its eigenvalue $\hat q^2-E^2$ and rescaling the
operators ${\cal K}_{i}$. One obtains thus the same dynamical
algebras as those governing the scalar modes but in different
representations \cite{CV4}.

The Casimir operators of the open algebra (\ref{algJK}) are
\begin{equation}\label{Cas}
{\cal C}_1= \vec{\cal J}^2{\cal B }^2 + \vec{\cal K}^2\,,\quad {\cal C}_2=
\vec{\cal J}\cdot \vec{\cal K}\,.
\end{equation}
In addition, we can define a new Casimir-type operator
\begin{equation}\label{4QI}
Q=\frac{\mu}{2}\,Q^Y + ({\cal B}-P_4) {\cal D}(\sigma_L+{\bf 1}_2)= {\cal
D}[\sigma_K+(\sigma_L+{\bf 1}_2)B]\,,
\end{equation}
which is an operator from ${\bf D}_0$ related to $Q^Y$. This satisfies the
simple algebraic relations,
\begin{equation}\label{QQKJ}
[Q, {\cal J}_i]=0\,, \quad [Q, {\cal K}_i]=0\,,\quad \{ Q, Q_i \}=2({\cal K}_i
+{\cal J}_i{\cal B})\,,
\end{equation}
and  the identity
\begin{equation}\label{Q2}
Q^2=\frac{\mu^2}{4}\,({P_4}I-{\cal B})^4 \,,
\end{equation}
resulting from Eqs. (\ref{C1C2}), (\ref{C1C21}) and (\ref{4QI}). Moreover,
using Eqs. (\ref{CBCB}) we find two new operators that can be put in a closed
form,
\begin{eqnarray}\label{CCC}
&&{\cal C}^+={\cal C}_1+2{\cal B}{\cal C}_2+{\cal B}^2=(Q+{\cal B})^2\,,\\
&&{\cal C}^-={\cal C}_1-2{\cal B}{\cal C}_2+{\cal B}^2=\frac{\mu^2}{4}\,(P_4
I+{\cal B})^4 \,.
\end{eqnarray}
The operators $Q$ and ${\cal C}^+$ are  Casimir operators only for the algebra
(\ref{algJK}) but $Q^2$ and ${\cal C}^-$ are general Casimir operators since
they commute with any other conserved Dirac operator.

Finally we observe that we can take over the operator $N$ of the scalar theory
since the Dirac and the Klein-Gordon particles have the same energy spectrum.
This offers us the opportunity  to introduce the new Casimir operator
\begin{equation}\label{MNP}
M=(N+\mu P_4)^2 I \,\in {\bf D}_0
\end{equation}
 that allows us to write $Q^2={\cal B}^2 M $, as
it results from Eqs. (\ref{CBCB}) and (\ref{Q2}).

\section{Infinite-dimensional  superalgebra}

Now we may ask how could be organized this very rich set of conserved Dirac
operators. There are many commutation and anticommutation relations that can
not be ignored such that it seems that the suitable structure may be a
superalgebra.  Thus we start with the open superalgebra ${\cal S}_0$ generated
by the operators $\{I, M, {\cal J}_i, {\cal K}_i, Q, Q_i\}\subset {\bf D}_0$
which satisfy Eqs. (\ref{algJK}), (\ref{QJK}) and (\ref{QQKJ}) completed with
the obvious anticommutation rule
\begin{equation}\label{QQM}
\{ Q,\,Q\}=2{\cal B}^2 M \,.
\end{equation}
We observe that, as in the non-relativistic quantum Kepler problem, there are
algebraic relations which remain open because of the factors ${\cal B}$.
Therefore, we are forced  to embed all the above ingredients in an {\em
infinite-di\-men\-sional} superalgebra constructed  in the same manner as the
infinite algebra of Ref. \cite{DD}. The difference is that here we  have a
superalgebra with generators of bosonic or fermionic type.

Let us define of the {\em bosonic} operators
\begin{equation}\label{defEJK}
I_n=I{\cal B}^n\,,\quad M_n=M{\cal B}^n  \,, \quad J_n^i={\cal J}_i{\cal
B}^n\,,\quad K_n^i={\cal K}_i{\cal B}^n\,,
\end{equation}
and the supercharges of the {\em fermionic} sector
\begin{equation}\label{defQQ}
Q_n=Q {\cal B}^n\,,\qquad Q_n^i={Q}_i{\cal B}^n \,,
\end{equation}
for any $n= 0,1,2...$.  The operators $I_n$ and $M_n$ are Casimir-type
operators commuting between themselves and with all the operators of the
bosonic or fermionic sectors. Then, according to Eqs. (\ref{algJK}) and
(\ref{defEJK}), we obtain the following non-trivial commutators of the bosonic
sector
\begin{eqnarray}
\left[ J^i_n, J^j_m \right] &=& i\varepsilon_{ijk}J^{k}_{n+m}\,,\\
\left[ J^{i}_n, K^{j}_m \right] &=& i\varepsilon_{ijk}K^k_{n+m}\,,\\
\left[ K^i_n, K^j_m \right] &=& i\varepsilon_{ijk}J^k_{n+m+2}\,,\label{sic}
\end{eqnarray}
while from Eqs. (\ref{Pau}),(\ref{QQKJ}) and (\ref{QQM})  we deduce the
anticommutators of the fermionic sector,
\begin{eqnarray}
\{ Q^i_n,Q^j_m \}&=&2\delta_{ij}I_{n+m}\,,\\
\{ Q_n, Q^i_m \}&=&2( K^i_{n+m} + J^i_{n+m+1})\,,\label{cuc}\\
\{Q_n, Q_m \}&=& 2\, M_{m+n+2}\,.\label{muc}
\end{eqnarray}
The commutations relations between the bosonic and fermionic operators are
\begin{eqnarray}
\left[ Q_n,J^j_m \right]=0 \,,&\quad&
[ Q^i_n,J^j_m ]=i\varepsilon_{ijk}Q^k_{n+m}\,,\\
\left[ Q_n,K^j_m \right]=0\,,&\quad& [ Q^i_n,K^j_m
]=i\varepsilon_{ijk}Q^k_{n+m+1}\,.\label{cucu}
\end{eqnarray}

Thus we constructed an infinite-dimensional superalgebra  ${\cal S}$ generated
by the countable set of operators $\{I_n,M_n,J^i_n,K^i_n,Q_n,Q^i_n\}$, $n\ge
0$. We observe that the typical algebraic structure related to the Kepler
problem is the infinite-dimensional algebra ${\cal A}$ generated by
$\{J^i_n,K^i_n\}$ which is a subalgebra in ${\cal S}$.

\section{Twisted loop superalgebras}

Now we intend to show  that the superalgebra ${\cal S}$ can be seen as a {\em
twisted} Kac-Moody superalgebra such that its subalgebra ${\cal A}$ should be a
twisted loop algebra of the usual $so(4)$ algebra, in the sense of Ref.
\cite{DD}.

First we define the finite-dimensional superalgebra ${\cal W}_0$ generated by
the operators $\{ E, F, A^i, B^i,G, G^i\}$. We assume that $E$ and $F$ commute
with any other generator and that the generators $\{A^i,B^i\}$ satisfy the
$so(4)$ algebra,
\begin{equation}\label{algAB}
[ {A}^{i},\, {A}^{j}]=i\varepsilon _{ijk}{A}^{k} \,,\quad [ {A}^{i},\,
{B}^{j}]=i\varepsilon _{ijk}{B}^{k} \,,\quad [ {B}^{i},\, {B}^{j}]=i\varepsilon
_{ijk}{A}^{k}\,.
\end{equation}
The operators $G$ and $G^i$ are fermionic supercharges obeying
\begin{equation}\label{algAB1}
\{ G^i, G^j\}=2\delta_{ij}E\,, \quad \{ G, G^i\}=2\,(A^i+B^i)\,,\quad\left\{ G,
G\right\}=2\,F
\end{equation}
and the commutation relations
\begin{eqnarray}
 [ A^{i},\, G ]=0\,,&\quad &[ {A}^{i},\, {G}^{j} ]=i\varepsilon _{ijk}{G}^{k} \,,\nonumber\\
 {[ B^{i},\, G ]}=0\,,&\quad &[ {B}^{i},\, {G}^{j} ]=i\varepsilon
 _{ijk}{G}^{k}\,.\label{algAB2}
\end{eqnarray}
This superalgebra has simple finite-dimensional representations as we briefly
present in Appendix.

Furthermore, we consider the corresponding Kac-Moody infinite loop superalgebra
${\cal W}$ generated by the operators $\{ E_n, F_n, A_n^i, B_n^i, G_n,
G_n^i\}$, $n\in {\Bbb Z}$, with the following properties
\begin{eqnarray}
\left[ {A}^{i}_n,\, {A}^{j}_m\right]&=&i\varepsilon_{ijk}{A}^{k}_{n+m}
\,,\qquad \left\{ G^i_n, G^j_m\right\}=2\delta_{ij}E_{n+m}\,,\nonumber\\
\left[ {A}^{i}_n,\, {B}^{j}_m\right]&=&i\varepsilon_{ijk}{B}^{k}_{n+m}\,,
\qquad \left\{ {G}_n,\, {G}^{j}_m\right\}=2\,(A^i_{n+m}+B^i_{n+m})\,,\label{algABn}\\
\left[ {B}^{i}_n,\, {B}^{j}_m\right]&=&i\varepsilon_{ijk}{A}^{k}_{n+m}\,,
\qquad \left\{ {G}_n,\, {G}_m\right\}=2\, {F}_{n+m}\,,\nonumber
\end{eqnarray}
and
\begin{eqnarray}
 [ A^{i}_n,\, G_m ]=0\,,&\quad &[ {A}^{i}_n,\, {G}^{j}_m ]=i\varepsilon _{ijk}{G}^{k}_{n+m} \,,\nonumber\\
 {[ B^{i}_n,\, G_m ]}=0\,,&\quad &[ {B}^{i}_n,\, {G}^{j}_m ]=i\varepsilon
_{ijk}{G}^{k}_{n+m}\,,
\end{eqnarray}
understanding that the generators $E_n$ and $F_n$, $n\in {\Bbb Z}$, commute
with any other generator of ${\cal W}$.

The next step is to define the involution automorphism $\tau : {\cal W}\to
{\cal W}^{\tau}$ selecting the countable subset of operators
\begin{equation}\label{setop}
\{ E_{2n}, F_{2n},A_{2n}^i, B_{2n+2}^i, G_{2n+2}, G_{2n}^i\}\,,\quad n\in {\Bbb
Z} \,,
\end{equation}
which generates the superalgebra ${\cal W}^{\tau}\subset {\cal W}$. The
algebraic properties of this superalgebra are given by the commutation
relations of the bosonic sector,
\begin{eqnarray}
\left[ {A}^{i}_{2n},\, {A}^{j}_{2m}\right]&=&i\varepsilon_{ijk}{A}^{k}_{2(n+m)}\,,\nonumber\\
\left[ {A}^{i}_{2n},\, {B}^{j}_{2m+2}\right]&=&i\varepsilon_{ijk}{B}^{k}_{2(n+m)+2}\,,\label{algABns}\\
\left[ {B}^{i}_{2n+2},\,
{B}^{j}_{2m+2}\right]&=&i\varepsilon_{ijk}{A}^{k}_{2(n+m+2)}\,,\nonumber
\end{eqnarray}
the anticommutation relations of the fermionic sector
\begin{eqnarray}
\left\{G^i_{2n}, G^j_{2m}\right\}&=&2\delta_{ij}E_{2(n+m)}\,,\nonumber\\
\left\{G_{2n+2}, G^j_{2m}\right\}&=&2\,(A_{2(n+m+1)}+B_{2(n+m)+2})\,,\label{algABns1}\\
\left\{G_{2n+2}, G_{2m+2}\right\}&=&2\,F_{2(n+m+2)} \,,\nonumber
\end{eqnarray}
and the commutation relations among both sectors,
\begin{eqnarray}
 [ A^{i}_{2n},\, G_{2m+2} ]=0\,,&\quad &[ {A}^{i}_{2n},\, {G}^{j}_{2m} ]=i\varepsilon _{ijk}{C}^{k}_{2(n+m)}
\,,\nonumber\\
 {[ B^{i}_{2n+2},\, G_{2m+2} ]}=0\,,&\quad &[ {B}^{i}_{2n+2},\, {G}^{j}_{2m} ]=i\varepsilon _{ijk}{C}^{k}_{2(n+m+1)}
\,.\label{algABns2}
\end{eqnarray}
In this way we have constructed the {\em twisted} loop superalgebra ${\cal
W}^{\tau}$ the positive part of which (with $n\ge 0$)  will be denoted by
${\cal W}^{\tau}_+$.

Now we can show  that the mapping $\phi\, :\, {\cal W}^{\tau}_+ \to {\cal S}$
defined by
\begin{equation}\label{IJKQ}
I_n=\phi(E_{2n})\,,\quad M_n=\phi(F_{2n})\,,\quad J^i_n=\phi(A^i_{2n})\,,\quad
K^i_n=\phi(B^i_{2n+2})\,,
\end{equation}
and
\begin{equation}\label{IJKQ1}
{Q}_n=\phi(G_{2n+2})\,,\qquad Q^i_n=\phi(G^i_{2n})\,,\quad n=0,1,2,...
\end{equation}
is an {\em homomorphism}. Indeed, if we consider, for example, the last of Eqs.
(\ref{sic}) we can write
\begin{eqnarray}
\left[\phi(B^i_{2n+2}),\phi(B^j_{2m+2}) \right]&=&\left[ K^i_n,
K^j_m \right]=i\varepsilon_{ijk}J^k_{n+m+2}\\
&=&i\varepsilon _{ijk}\phi({A}^{k}_{2(n+m+2)})
=\phi(\left[B^i_{2n+2},B^j_{2m+2} \right])\,.\nonumber
\end{eqnarray}
In this manner one can demonstrate step by step that for any pair of
generators, $X$ and $Y$, of ${\cal W}^{\tau}_+$ we have either $[\phi(X),
\phi(Y)]=\phi([X,Y])$ or $\{\phi(X), \phi(Y)\}=\phi(\{X,Y\})$. Reversely, if we
start with Eqs. (\ref{IJKQ}) and (\ref{IJKQ1}) supposing that $\phi$ is an
homomorphism, then we recover the superalgebra ${\cal S}$. For example, the
last of Eqs. (\ref{cucu}) results from
\begin{eqnarray}
\left[ Q^i_n, K^j_m \right]&=& \left[\phi(G^i_{2n}),\phi(B^j_{2m+2}) \right]
=\phi(\left[G^i_{2n},B^j_{2m+2} \right])\nonumber\\
 &=&i\varepsilon
_{ijk}\phi({G}^{k}_{2(n+m+1)}) =i\varepsilon_{ijk}Q^k_{n+m+1}\,.
\end{eqnarray}
In this way, we bring arguments that our superalgebra ${\cal S}$ can be seen as
a twisted loop superalgebra.

It finally should be mentioned that the above construction of the twisted loop
superalgebras could be regarded differently. The connection between the set of
operators (\ref{defEJK}), (\ref{defQQ}) and (\ref{setop}) can be realized
directly assigning grades to each operator \cite{daboul} as follows:
\begin{eqnarray}
&&E_{2n}:=I{\cal B}^n\,,\quad F_{2n}:=M{\cal B}^n  \,,
\quad A_{2n}^i:={\cal J}_i{\cal B}^n\,,\nonumber\\
&&B_{2n+2}^i:={\cal K}_i{\cal B}^n\,,\quad G_{2n+2}:=Q {\cal B}^n\,,\quad
G_{2n}^i:={Q}_i{\cal B}^n \,.
\end{eqnarray}
Thus we achieve a graded loop superalgebra of the Kac-Moody type and the sum of
the grades is conserved under (anti)commutations.

\section{Concluding remarks}

Here we constructed the infinite-dimensional superalgebra ${\cal S}$ starting
with the finite-dimensional open superalgebra ${\cal S}_0$ formed only by
conserved operators commuting with $I$, $H$, $P_4$ and the whole set of Casimir
operators freely generated by these three operators.

In ${\cal S}_0$ we explicitly used two Casimir operators, namely $I$ and $M$.
As mentioned, $I$ is the projector on the physical spinor subspace playing the
role of identity operator. More interesting is the operator $M$ since this
depends on $N$ which is in some sense similar with the operator
$(-2H_K)^{-1/2}$ of the $so(4,2)$ dynamical algebra of the quantum Kepler
problem governed by the non-relativistic Hamiltonian operator
$H_K=-\frac{1}{2}\Delta - r^{-1}$ \cite{BARUT}. We remind the reader that the
$so(4,2)$ dynamical algebra of the Kepler problem contains not only conserved
operators but even operators that do not commute with $H_K$. In this case, the
conserved operators, i. e. the angular momentum and the Runge-Lenz vector
operator, are {\em orthogonal}  and generate an {\em open} algebra that can be
rescaled obtaining thus the dynamical algebra $o(4)\subset so(4,2)$ of the
discrete energy spectrum. The first Casimir operator of $o(4)$, that reads
$C_K^1=(-2H_K)^{-1}-I$, has a similar form with our operator $M$. However, the
second Casimir operator of $o(4)$ vanishes while our operator ${\cal C}_2$,
given by Eq. (\ref{Cas}), is different from zero since the vector operators
$\vec{\cal J}$ and $\vec{\cal K}$ are not orthogonal.

In these circumstances we can say that the subalgebra (\ref{algJK}) of ${\cal
S}_0$ corresponds to the open algebra that gives the $o(4)$ dynamical algebra
of the Kepler problem. This explains why our twisted loop superalgebra was
constructed in a similar way as that of the Kepler case \cite{DD}. The main
difference between these two theories is that ${\cal S}_0$ is an open
superalgebra containing the supercharges $Q$ and $Q_i$ that naturally arise
from the very special geometry of the Euclidean Taub-NUT space. For this reason
we were forced to include, in addition,  the bosonic Casimir operator $M$ for
writing down Eq. (\ref{QQM}). We specify that this is more than a simple
artifice since the resulting infinite superalgebra ${\cal S}$ is a twisted loop
superalgebra  arising from  a coherent algebraic structure, namely the
superalgebra ${\cal W}_0$ the representations of which are presented in
Appendix.

In other respects, it is clear that the operators $Q$ and $Q_i$ appear only in
the Dirac theory on ${\frak M}$ since they are in fact Dirac-type operators.
Therefore, it is interesting to compare our results with relativistic systems
with spin half whose non-relativistic limit is the quantum Kepler problem.
Thus, in the case of the Dirac electron in external Coulomb field there exists
a hidden symmetry even if one does not have a conserved Runge-Lenz operator.
This symmetry is related to another operator, called the Johnson-Lippmann
operator \cite{JL}, that is a scalar conserved operator. In the
non-relativistic limit this becomes the projection of the usual Runge-Lenz
vector operator of the Kepler problem on the electron spin direction \cite{KK}.
In our approach, we can say that the Johnson-Lippmann operator is just the
supercharge $Q$ whose first term  given by Eq. (\ref{4QI}) is ${\cal
D}(\vec{\sigma}\cdot \vec{K})$.

Finally we note that our open superalgebra ${\cal S}_0$ could be enlarged
adding non-conserved operators that can be either leader operators or operators
related to the manifest supersymmetry \cite{CV2} of our Hamiltonian $H$.
However, this problem will be considered elsewhere.

\subsection*{Acknowledgments}

We are grateful to Laszlo Feher for interesting and useful discussions
on closely related subjects. The work was supported in part by CNCSIS grant 
1154/2007 and CEEX program CEx 05-D11-49.

\setcounter{equation}{0} \renewcommand{\theequation}
{A.\arabic{equation}}

\section*{Appendix \\
The superalgebra ${\cal W}_0$}

Here we would like to show that a fundamental representation of the
superalgebra ${\cal W}_0$ arises from a particular representation of the
$so(4)$ Lie algebra.

We start with a finite-dimensional representation, $\rho$, of this algebra
generated by the linear operators $\{A^i_{\rho},B^i_{\rho}\}$ defined on the
space ${\frak M}_{\rho}$ and obeying Eqs. (\ref{algAB}). The identity operator
of on ${\frak M}_{\rho}$ is denoted by $1_{\rho}$.  The $su(2)\times su(2)$
content of the $so(4)$ algebra can be pointed out in the new basis
$\{J^{i}_{\rho\,+},J^{i}_{\rho\,-}\}$ given by the operators
$J^{i}_{\rho\,\pm}= \frac{1}{2}(A_{\rho}^i\pm B_{\rho}^i)$ that satisfy the
$su(2)$ commutation relations,
\begin{equation}\label{algJJ}
[ {J}^{i}_{\rho\,+},\, {J}^{j}_{\rho\,+}]=i\varepsilon _{ijk}{J}^{k}_{\rho\,+}
\,,\quad [ {J}^{i}_{\rho\,-},\, {J}^{j}_{\rho\,-}]=i\varepsilon
_{ijk}{J}^{k}_{\rho\,-} \,,\quad [ {J}^{i}_{\rho\,+},\, {J}^{j}_{\rho\,-}]=0\,.
\end{equation}
The representation $\rho=(j^+,j^-)$ is completely determined by the $su(2)$
weights defined by the Casimir operators
$\vec{J}^2_{\rho\,\pm}=j^{\pm}(j^{\pm}+1)\,1_{\rho}$. However, here we have to
consider, in addition, the usual Casimir operators
$C_{\rho\,1}=\vec{A}^2_{\rho}+\vec{B}^2_{\rho}$ and
$C_{\rho\,2}=\vec{A}_{\rho}\cdot\vec{B}_{\rho}$ or the new ones
\begin{equation}
C_{\rho \pm}=C_{\rho\, 1}\pm 2 C_{\rho\, 2} +1_{\rho}=4 \vec{J}_{\rho\, \pm}^ 2
+ 1_{\rho} = (2j^{\pm}+1)^2\, 1_{\rho} \,.
\end{equation}
We note that when $j^+=j^-$ then $\vec{A}_{\rho}$ and $\vec{B}_{\rho}$ are
orthogonal, as in the case of the dynamical algebra of the quantum Kepler
problem.

Our purpose is to construct the superalgebra ${\cal W}_0$ in the carrier space
${\frak M}={\frak M}_{\rho}\otimes {\frak M}_{(\frac{1}{2},0)}$ of the
reducible representation $(j^+,j^-)\otimes (\frac{1}{2},0)$ given by arbitrary
weights $j^{\pm}$ taking positive real values. The representation,
$(\frac{1}{2},0)$, is generated by the operators $\hat A^i=\frac{1}{2}\sigma_i$
and  $\hat B^i=\frac{1}{2}\sigma_i$ acting in the two-dimensional space ${\frak
M}_{(\frac{1}{2},0)}$ where the identity operator is $1_2$. In these
circumstances we define first the identity operator on ${\frak M}$,
$E=1_{\rho}\otimes 1_2$, and the Casimir-type operator $F=C_{\rho\,+}\otimes
1_2$. The $so(4)$ generators of this representation are
\begin{equation}
{A}^i=A^i_{\rho}\otimes 1_2 +{\textstyle \frac{1}{2}}1_{\rho} \otimes
\sigma_i\,,\quad {B}^i=B^i_{\rho}\otimes 1_2 +{\textstyle
\frac{1}{2}}1_{\rho}\otimes \sigma_i\,.
\end{equation}
Moreover, we introduce the supercharges
\begin{equation}
G_i=1_{\rho}\otimes \sigma_i\,, \quad G=\vec{A}_{\rho}\otimes
\vec{\sigma}+\vec{B}_{\rho}\otimes \vec{\sigma}+E\,,
\end{equation}
so that $G^2=F$. Now it is a simple exercise to show that the operators $\{E,F,
A^i,B^i,G,G^i\}$ satisfy Eqs. (\ref{algAB}), (\ref{algAB1}) and (\ref{algAB2}).

The conclusion is that the superalgebra ${\cal W}_0$ can be realized in the
carrier space of any reducible representation $\rho\otimes (\frac{1}{2},0)$ of
the $so(4)$ algebra.

\end{document}